\documentclass[12pt]{article}
\usepackage{graphicx}
\usepackage{cite}
\usepackage{amssymb}

\newcommand\pubnumber{}
\newcommand\pubdate{\today}

\def\ific{Instituto de F\'isica Corpuscular - IFIC\\
CSIC-UVEG, Apartado de Correos 22085, E-46071 Valencia, SPAIN}
\def\support{}

\def\Title#1{\begin{center} {\Large #1 } \end{center}}
\def\Author#1{\begin{center}{ \sc #1} \end{center}}
\def\Address#1{\begin{center}{ \it #1} \end{center}}

\newcommand\pubblock{\rightline{\begin{tabular}{l} \pubnumber\\
         \pubdate  \end{tabular}}}
\newenvironment{Abstract}{\begin{quotation}  }{\end{quotation}}
\newenvironment{Presented}{\begin{quotation} \begin{center} 
             PRESENTED AT\end{center}\bigskip 
      \begin{center}\begin{large}}{\end{large}\end{center} \end{quotation}}
\def\Acknowledgements{\bigskip  \bigskip \begin{center} \begin{large}
             \bf ACKNOWLEDGEMENTS \end{large}\end{center}}

\def\beq{\begin{equation}}
\def\eeq#1{\label{#1}\end{equation}}
\def\eeqn{\end{equation}}

\def\beqa{\begin{eqnarray}}
\def\eeqa#1{\label{#1}\end{eqnarray}}
\def\eeqan{\end{eqnarray}}

\let\bar=\overbar

\def\Dslash{\not{\hbox{\kern-4pt $D$}}}
\def\dslash{\not{\hbox{\kern-2pt $\del$}}}

\def\msb{{\bar{\ssstyle M \kern -1pt S}}}

\newcommand{\athdm}[0]{A2HDM }
\newcommand{\athdmws}[0]{A2HDM}
\newcommand{\thdm}[0]{2HDM }
\newcommand{\thdmws}[0]{2HDM}

\begin{document}
\begin{titlepage}
\pubblock

\vfill
\Title{Flavour Physics in two-Higgs-doublet models}
\vfill
\Author{ Martin Jung\support}
\Address{\ific}
\vfill
\begin{Abstract}
Despite the tremendous success of the Standard Model, the arguments for the necessity of an extension are compelling. An attractive option is provided by Two-Higgs-Doublet models, 
due to their simplicity and them being the low-energy limit of some more complete theories. In the most general version of the model, the fermionic couplings of the neutral scalars are non-diagonal in flavour and, therefore, generate unwanted flavour-changing neutral-current phenomena. Different ways
to suppress FCNCs have been developed, giving rise to a variety of specific implementations of the \thdmws. Three of these are discussed in this talk, comparing their phenomenological influence in flavour observables: The use of a discrete $\mathcal{Z}_2$-symmetry, an expansion around this limit in a minimal flavour violation scenario assuming the decoupling limit, and the Aligned Two-Higgs-Doublet Model. All of these result in the absence of tree-level flavour-changing neutral currents. Their different phenomenological consequences are demonstrated for a selection of observables, namely (semi-) leptonic decays, $\bar{B}\to X_s\gamma$ and mixing in the $B^0_{d,s}$-systems.
\end{Abstract}
\vfill
\begin{Presented}
CKM2010 -- 6th International Workshop on the CKM Unitarity Triangle\\
Warwick, United Kingdom,  September 6--10, 2010
\end{Presented}
\vfill
\end{titlepage}
\def\thefootnote{\fnsymbol{footnote}}
\setcounter{footnote}{0}

\section{Introduction}
The confirmation by the $B$-factories of the CKM-mechanism as the main source for low-energy $CP$-violation has opened a new chapter in the search for physics beyond the Standard Model (SM). While the direct search for the mediator(s) of electroweak symmetry breaking and new heavy particles is being performed at the Tevatron and the LHC, their effects should be visible in (low-energy) flavour observables as well. Loosely speaking, the infamous \emph{flavour problem} is the fact, that this is not the case (yet), implying a highly non-trivial flavour-structure of any new physics (NP) scenario. While not conclusive, some tensions are present in flavour-data, two of which will be discussed here: the different values for $\sin2\beta$ when extracted with the help of $B\to J/\psi K$ and $B\to\tau\nu$ \cite{Deschamps:2008de,Bona:2009cj}, and hints for a large phase in $B_s$-mixing, especially the measurement of the sign-like dimuon asymmetry by D0 \cite{Abazov:2010hv}.

Regarding two-Higgs-doublet models (\thdmws) \cite{Lee:1973iz}, the many models in the literature (for references, see e.g. \cite{Gunion:1989we}) basically differ by the mechanism to avoid this problem, which renders their most general version unplausible. In this talk, the focus lies on three mechanisms all of which avoid flavour-changing neutral currents (FCNC) at tree-level: the use of a discrete $\mathcal{Z}_2$-symmetry \cite{Glashow:1976nt}, an expansion around this limit in a minimal flavour violation scenario assuming the decoupling limit \cite{Buras:2010mh,Buras:2010zm}, and the Aligned Two-Higgs-Doublet Model (\athdmws) \cite{Pich:2009sp,Jung:2010ik}. They are introduced in the next section, before their phenomenology is discussed in section~3 by examining selected flavour observables, followed by the conclusion in section~4.

\section{Models}
The quark Yukawa sector of the most general \thdm is given by
\begin{equation}
-\mathcal L_Y^q  \;=\; \bar{Q}_L' (\Gamma_1 \phi_1 +\Gamma_2 \phi_2)\, d_R' + \bar{Q}_L' (\Delta_1 \widetilde{\phi}_1 +\Delta_2 \widetilde{\phi}_2)\, u_R'
 + \mathrm{h.c.}\,,
\end{equation}
with $\Gamma_i,\Delta_i$ being $F\times F$ matrices, where $F$ denotes the number of families, and otherwise common notation (for details here and in the following see \cite{Jung:2010ik}). In models with a $\mathcal{Z}_2$-symmetry, each field gets assigned an additional parity-like quantum number, effectively forbidding one of the two possible couplings between identical fermion fields\footnote{The symmetry holds also for quantum corrections. However, once the \thdm is embedded into another theory providing a UV completion, typically quantum corrections break it, and potentially too large FCNCs are generated \cite{Buras:2010mh}. Here the \thdm without UV-completion is considered}. Recent phenomenological analyses include \cite{WahabElKaffas:2007xd,Deschamps:2009rh,Flacher:2008zq,Bona:2009cj,Mahmoudi:2009zx}. Minimal flavour violation (MFV), defined according to \cite{D'Ambrosio:2002ex}, is an effective field theory framework, in which the flavour symmetry of the SM Lagrangian without quark Yukawa couplings, namely $\mathcal{G}_F=SU(3)_L\times SU(3)_{U_R}\times SU(3)_{D_R}$\footnote{The discussion of the additional $U(1)$ symmetries is omitted here.}, is extended to the full Lagrangian by promoting the Yukawa couplings to spurion fields, transforming accordingly under $\mathcal{G}_F$. In addition it is assumed, that the CKM phase is the only source of $CP$ violation. This constrains heavily the structure of higher-dimensional operators, suppressing FCNCs effectively by powers of (light) quark masses and CKM mixing angles. In \cite{D'Ambrosio:2002ex} and \cite{Buras:2010mh,Buras:2010zm}, this program is carried out as an expansion around the limit of a \thdm Type II, relevant for SUSY, and assuming the decoupling limit $M_{H^\pm}\gg M_W$, resumming $\tan \beta$-enhanced terms. In the latter papers, the restriction regarding new $CP$-violating sources has been dropped\footnote{Regarding MFV with new sources of $CP$ violation, see also \cite{Colangelo:2008qp,Kagan:2009bn,Botella:2009pq,Paradisi:2009ey,Feldmann:2009dc}.}. The decoupling limit implies, that charged Higgs effects are usually negligible. In the \athdm it is assumed that each pair of coupling matrices is aligned, $\Delta_1\sim\Delta_2,\Gamma_1\sim\Gamma_2$, leading to
\begin{equation}\label{lagrangian}
\mathcal L_{Y,H^\pm}^{\rm{q}} = - \frac{\sqrt{2}}{v}\, H^+(x) \left\{  \bar{u}(x)   \left[  \varsigma_d\, V_{CKM} M_d \mathcal P_R - \varsigma_u\, M_u^\dagger V_{CKM} \mathcal P_L  \right]   d(x)\right\}+ \;\mathrm{h.c.} 
\end{equation}
for the charged scalar Yukawa interaction and the absence of FCNCs at tree level. The universal (flavour-blind) couplings $\varsigma_f$ ($f=u,d,l$) introduce three new complex phases
and, therefore, a new source of $CP$ violation. For particular (real) values of these parameters
the usual $CP$-conserving models based on discrete $\mathcal{Z}_2$ symmetries
are recovered.
Quantum corrections induce a misalignment of the Yukawa matrices, generating small
FCNC effects \cite{Pich:2009sp,Jung:2010ik,Ferreira:2010xe,Braeuninger:2010td}. The flavour symmetries of the A2HDM strongly constrain the allowed FCNC
structures, providing at the quantum level 
an explicit implementation of the MFV scenario,
but allowing at the same time for new $CP$-violating phases.



\section{Phenomenology}

The leptonic decay rate is modified in the presence of a charged scalar as $\frac{\Gamma(P^+_{ij}\rightarrow l^+ \nu_l)_{\rm full}}{\Gamma(P^+_{ij}\rightarrow l^+ \nu_l)_{\rm SM}}\, =\, |1-\Delta_{ij}|^2$, where $i,j$ represent the valence quarks of the meson $P$ under consideration and $\Delta_{ij}$ encodes model-dependent information about the charged Higgs couplings. In the \athdm it is given by $\Delta_{ij}\, =\,\left( \frac{m_{P^{\pm}_{ij}}}{M_{H^{\pm}}} \right)^2\varsigma_l^*\,\frac{\varsigma_um_{u_i}+\varsigma_dm_{d_j}}{m_{u_i}+m_{d_j}}$. Obviously, in the decoupling limit the large charged Higgs mass renders the influence tiny, allowing to use $B\to\tau\nu$ in the unitarity triangle (UT) fit. In $\mathcal{Z}_2$-models, the two contributions are related and have fixed signs, allowing e.g. in Type II models only for a reduction of the rate in $B\to\tau\nu$ \footnote{A huge NP contribution $\Delta_{ub}>2$ could lead to an enhancement, but is ruled out by other observables.}, while in the \athdm the two contributions are independent and their relative influence is determined by the corresponding phases.

Semileptonic decays receive contributions from a charged scalar as well, but in this case the leading SM amplitude is not helicity suppressed, therefore the relative influence is smaller.
In addition, there are two momentum-dependent form factors involved. 
The scalar-exchange amplitude only contributes to the scalar form factor; it amounts to a multiplicative correction
$
\tilde f_0(t)\, = \, f_0(t)\;\left( 1 + \delta_{ij}\, t\right)\, ,
$
where the $\delta_{ij}$ is the analogue to $\Delta_{ij}$ in leptonic decays discussed above, in the \athdm given by
$
\delta_{ij}\,\equiv \, -\frac{\varsigma_l^*}{M_{H^{\pm}}^2}\,
\frac{\varsigma_um_{u_i}-\varsigma_dm_{d_j}}{m_{u_i}-m_{d_j}}\,,
$
and the above qualitative discussion applies here as well.

\begin{figure}[htb]
\centering
\includegraphics[width=45mm]{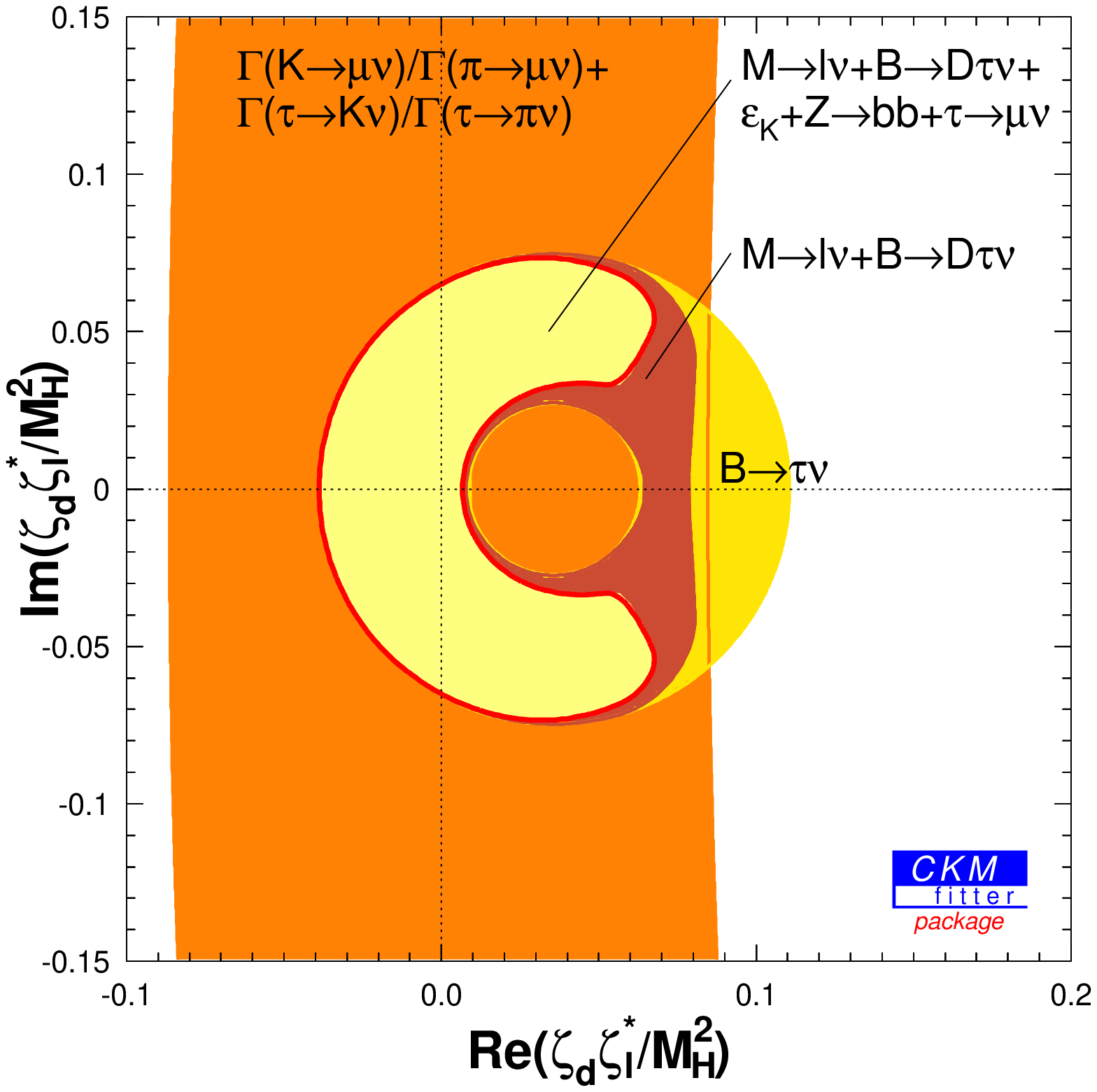}\quad\quad\includegraphics[width=45mm]{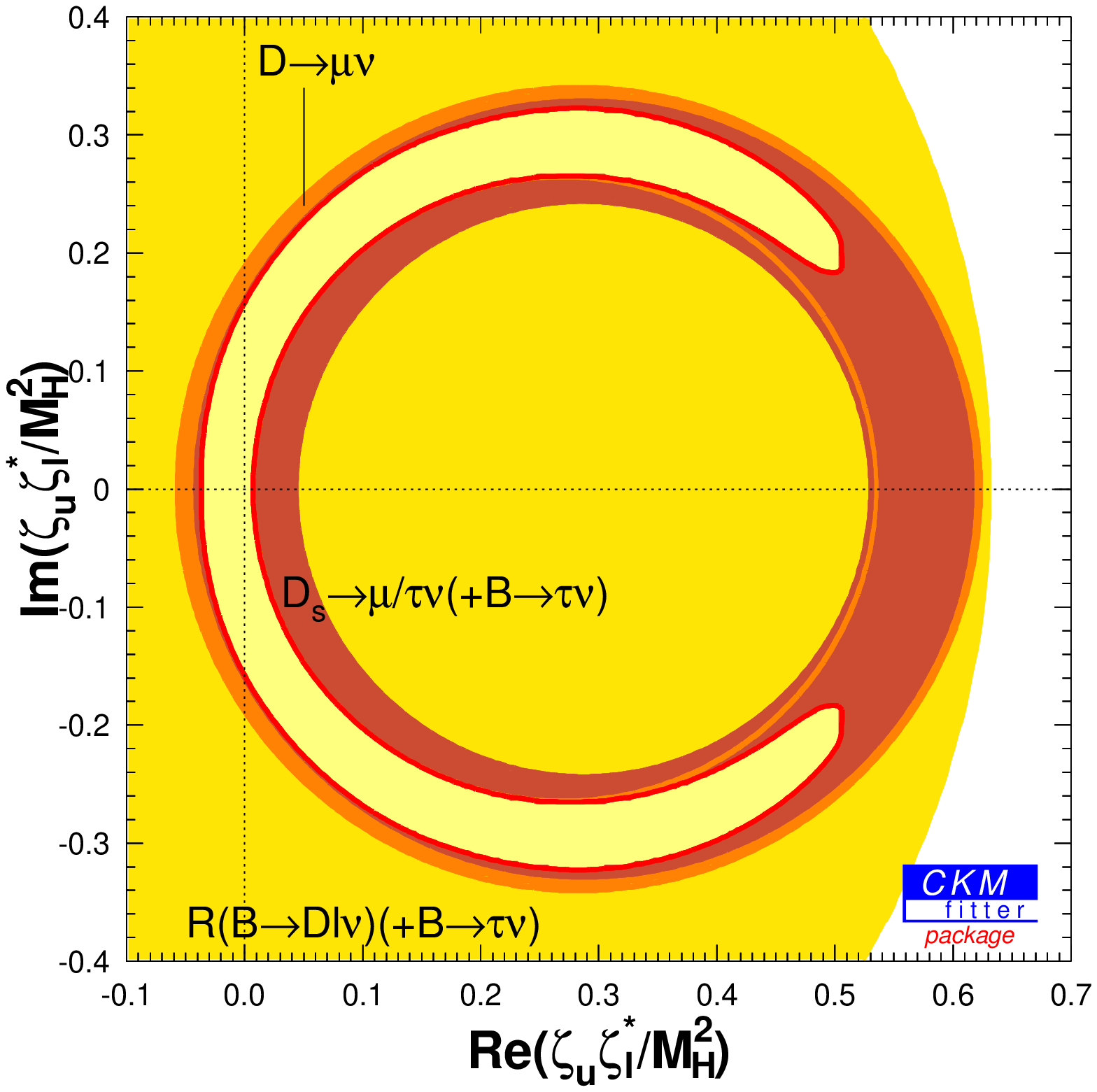}\quad\includegraphics[width=45mm]{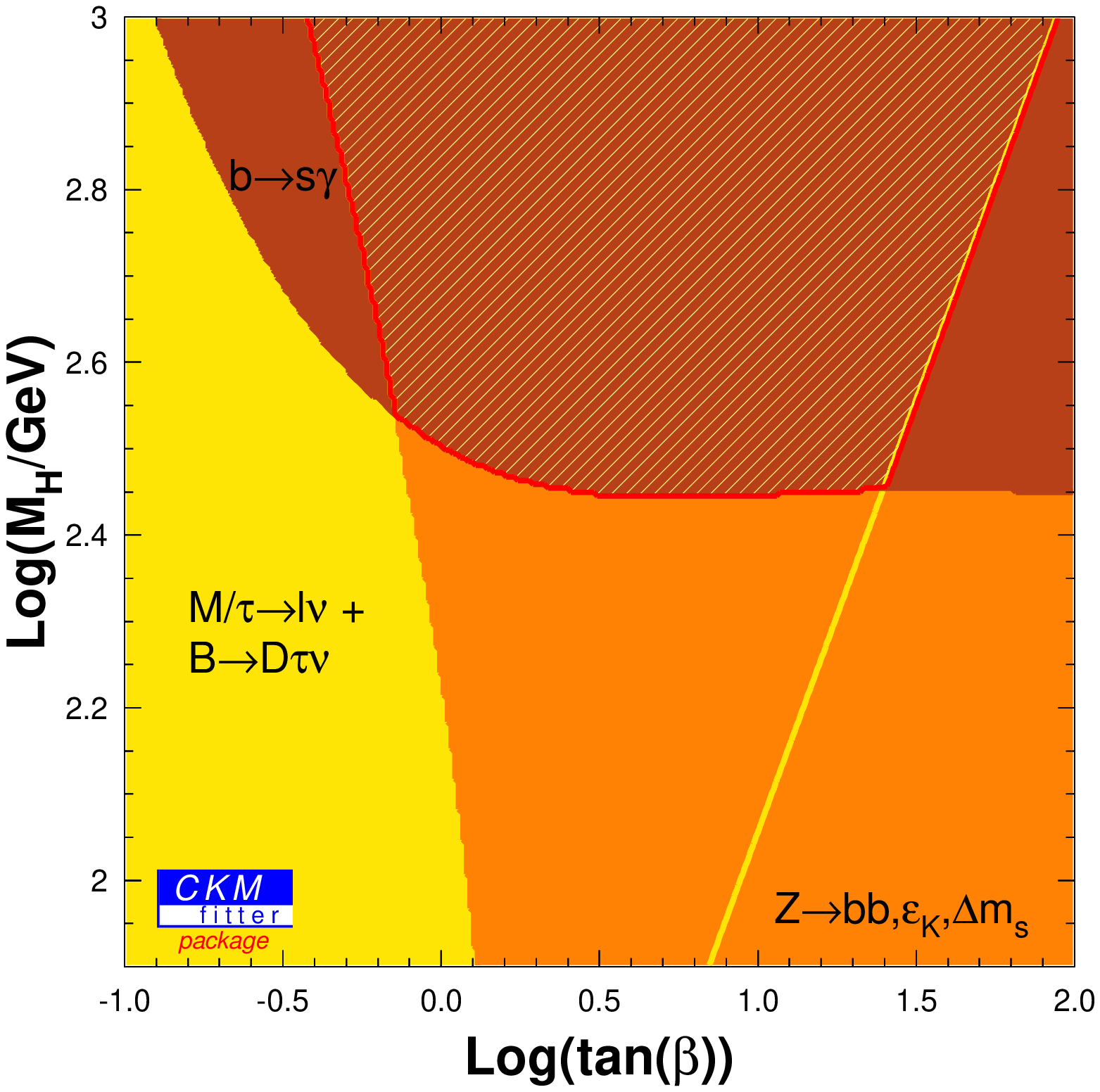}
\caption{\it $\varsigma_d\varsigma_l^*/M_{H^{\pm}}^2$ (left) and $\varsigma_u\varsigma_l^*/M_{H^{\pm}}^2$ (center) in the complex plane, in units of $GeV^{-2}$, constrained by leptonic and semileptonic decays. The inner yellow area shows the allowed region at 95\% CL, in the case of $\varsigma_d\varsigma_l^*/M_{H^{\pm}}^2$ using additional information \cite{Jung:2010ik}. Shown on the right is the projection to the Type II \thdm in the $\tan\beta-M_{H^\pm}$-plane, including information from loop-induced processes.}
\label{fig::globalsl}
\end{figure}

The results of a global fit to the available (semi-)leptonic decay data in the \athdm are shown in Fig.~\ref{fig::globalsl}.
Note that in both cases there are two real solutions, one of which can be excluded mainly due to the correlation provided by $B\to D\tau\nu$, but in the case of the combination $\varsigma_d\varsigma_l^*/M_{H^{\pm}}^2$ only with help of additional information \cite{Jung:2010ik}. For models with $\mathcal{Z}_2$-symmetry, only the projections on the real axes are relevant, the resulting constraint for the \thdm Type II is shown on the right.

Turning to loop-induced processes, maybe the most prominent example is the raditative decay $b\to s\gamma$, calculated basically up to NNLO in the SM \cite{Misiak:2006zs,Becher:2006pu}, see e.g. \cite{Hurth:2010tk} for recent developments and references. Combining high theoretical and experimental precision, it provides a very sensitive probe for FCNC effects. In the Type II \thdm it is famous for giving a bound on the Higgs mass basically independent of $\tan\beta$, as is illustrated in Fig.~\ref{fig::globalsl}, whose strength implies compatibility with the decoupling limit scenario. This cancellation is absent in general. In the \athdmws, the constraint on the single parameters is relatively weak as more parameters are involved \cite{Jung:2010ik}. However, the strength of the constraint appears now in the form of correlations with a large impact on related observables \cite{Jung:2010ab}.

Finally the effects in meson mixing are discussed. Models with $\mathcal{Z}_2$-symmetry do not effect these systems largely as long as $\tan\beta\gtrsim2$, which is generally assumed there. In the \athdm, the relevant coupling is independent, therefore effects from the top-coupling in $K$-mixing ($\epsilon_K$) and $B$-mixing are potentially large and constrain the parameter $|\zeta_u|\lesssim1$. They are universal in $B_{d,s}$, therefore the ratio $\Delta m_d/\Delta m_s$ can still be used in the UT fit. The mixing phase receives a moderate contribution from charged Higgs effects, up to $\sim 5$ times the SM value, taking into account the correlation from $b\to s\gamma$ \cite{Jung:2010ab}. In the decoupling scenario the dominant effects are from neutral Higgs exchange and proportional to down-type quark masses. Therefore the effect in $K$-mixing is tiny, while it is non-universal and potentially large in the $B_{d,s}$-systems \cite{Buras:2010mh}. The hint for a large NP phase in $B_s$-mixing can be accomodated here by a large phase, which implies a small shift in the $B_d$-system as well, in the right direction concerning the tension in the UT-fit.

\section{Conclusions}
\thdmws s remain an active field, providing a relatively simple extension of the SM with interesting influence on flavour observables. Models with $\mathcal{Z}_2$-symmetry are the best constrained, but do not offer new sources of $CP$-violation and might be problematic regarding their UV-completion \cite{Buras:2010mh}. The \athdm introduces new sources of $CP$ violation in the flavour sector while avoiding FCNCs at tree level, and provides an explicit counter-example to the widespread assumption that
in \thdmws s without tree-level-FCNCs all $CP$-violating phenomena should originate in the CKM matrix. Since all Yukawa couplings are proportional to fermion masses, the \athdm gives rise to an interesting hierarchy of FCNC effects, avoiding the stringent experimental constraints for light-quark systems and allowing at the same time for interesting signals in heavy-quark transitions. It affects all of the present tensions; however, a very large effect in $B_s$-mixing seems difficult to accomodate with charged-Higgs-effects only.
The focus of the decoupling MFV-scenario is very different: it describes radiatively induced corrections to the Type II model, providing a very different pattern of NP effects, capable of addressing the present tensions in the UT-fit and $B$-mixing. Note that these contributions are present in the \athdm as well. With the LHC up and running, and several experiments under planning and construction, prominently Super-$B$ factories, the experimental situation will greatly improve in the coming years, allowing a determination of the couplings in the different models discussed here, or their exclusion.
\vspace{-0.7cm}
\Acknowledgements
I thank the organizers of the workshop for the invitation, and Toni Pich and Paula Tuz\'on for a fruitful collaboration. This work has been supported in part 
by the EU MRTN network FLAVIAnet [Contract No. MRTN-CT-2006-035482], 
by MICINN, Spain 
[Grants FPA2007-60323 and Consolider-Ingenio 2010 Program CSD2007-00042 --CPAN--] and by Generalitat Valenciana [Prometeo/2008/069].

\end{document}